# Artificial Neural Network For Transient Stability Assessment: A Review

**Umair Shahzad**

Department of Electrical and Computer Engineering,
University of Nebraska-Lincoln,
Lincoln, NE, USA
umair.shahzad@huskers.unl.edu

*Abstract*–Integration of large-scale renewable energy sources and increasing uncertainty has drastically changed the dynamics of power system and has consequently brought various challenges. Rapid transient stability assessment of modern power system is a vital requirement for accurate power system planning and operation. The conventional methods are unable to fulfil this requirement. Therefore, novel approaches are required in this regard. Machine leaning approaches such as artificial neural network can play a significant role in this regard. Therefore, this paper aims to review the application of artificial neural network for transient stability assessment of power systems. It is believed that this work will provide a solid foundation for researchers in the domain of machine learning applications to power system security and stability.

TABLE I. TABLE OF ABBREVIATIONS

| Abbreviation | Meaning |
|---|---|
| ANN | Artificial Neural Network |
| CCT | Critical Clearing Time |
| CGAN | Conditional Generative Adversarial Network |
| CNN | Convolution Neural Network |
| DL | Deep Learning |
| DSA | Dynamic Security Assessment |
| DT | Decision Tree |
| EEAC | Extended Equal Area Criterion |
| FNN | Feedforward Neural Network |
| GCN | Graph Convolutional Network |
| GGNN | Gated Graph Neural Network |
| GIS | Geographic Information System |
| GNN | Graph Neural Network |
| HVDC | High Voltage Direct Current |
| KNN | K-Nearest Neighbor |
| LSTM | Long Short-Term Memory |
| ML | Machine Learning |
| MLPNN | Multi-layer Perceptron Neural Network |
| PMU | Phasor Measurement Unit |
| PNN | Probabilistic Neural Network |
| PTS | Probabilistic Transient Stability |
| RBF | Radial Basis Function |
| RGCN | Recurrent Graph Convolutional Network |
| RL | Reinforcement Learning |
| RNN | Recurrent Neural Network |
| SL | Supervised Learning |
| SML | Supervised Machine Learning |
| SVM | Support Vector Machine |
| TEF | Transient Energy Function |
| TSA | Transient Stability Assessment |
| UL | Unsupervised Learning |
| UML | Unsupervised Machine Learning |
| UPFC | Unified Power Flow Controller |

## I. INTRODUCTION

One of the main requirements for a reliable power system is to maintain the synchronous generators running in parallel and with sufficient capability to meet the load demand. In power system, the term transient stability is defined as the ability of the synchronous machines to remain in synchronism during the brief time following a large disturbance, such as a three-phase fault [1-2]. There are various conventional methods to assess transient stability; however, the most common one is the time-domain simulation approach. This is popular due to its universal nature and high accuracy; however, it is too time consuming which makes it unsuitable for real time prediction of transient stability [3]. The Transient Energy Function (TEF) method and Extended Equal Area Criterion (EEAC) have also been used to assess power system transient stability. Though, these approaches have some modeling restrictions, and they require numerous computations to compute transient stability index [4].

In view of the above mentioned, there is a dire need to explore new horizons in terms of assessing transient stability. Various computational intelligence-based machine learning (ML) approaches such as Artificial Neural Network (ANN), Support Vector Machine (SVM), and Decision Tree (DT) have been proposed till date for this purpose. Among these approaches, ANN is the most common and popular one due to various reasons. It does not require any rigorous mathematical modeling for its training, and it has a modular structure which allows parallel processing. Moreover, it has the ability to instantly map nonlinear relations between input data and output data [1]. Therefore, this paper will attempt to specifically review the application of ANN to Transient Stability Assessment (TSA).

The rest of the paper is organized as follows. Section II discusses background and overview of ML. Section III provides a classification of ML. Section IV provides background and overview of ANN. Section V briefly describes various components of ANN. Section VI provides a summary of various work of ANN application to TSA. Section VII provides research gaps and suggestions for future work. Finally, Section VIII concludes the paper.

## II. MACHINE LEARNING: OVERVIEW AND BACKGROUND

ML is broadly regarded as the subset of artificial intelligence [5] (simulation of human intelligence in machines, which are programmed to think like humans and mimic their actions), as outlined by Fig. 1. ML basically is an application of artificial intelligence that provides systems the ability to automatically learn and enhance from experience without being explicitly programmed [5-6]. In fact, the ML performs data analysis, using a set of instructions, through a variety of algorithms, for decision making and/or predictions [7]. Laborious designing and programming of algorithms are essential to be conducted, for ML, to implement diverse functionalities, such as, classification, clustering, and regression. Deep learning (DL) is a class of ML algorithms that uses multiple layers to progressively extract higher-level features from the raw input. For example, in image processing, lower layers may identify edges, whereas higher layers may identify the concepts relevant to a human being, such as digits, letters or faces [8]. It is majorly used for speech recognition, computer vision (high-level understanding from digital images or videos), medical image analysis, and natural language processing. There are several architectures used in DL such as deep neural networks, deep belief networks, recurrent neural networks, long short-term memory, and convolutional neural networks. The DL generally requires huge processing power and massive data [8]. The focus of this work is, however, on ML.

ML differs from traditional programming, in a very distinct manner. In traditional programming, the input data and a well written and tested program is fed into a machine to produce output. When it comes to ML, input data along with the output is fed into the machine during the learning phase, and it works out a program for itself. This is illustrated in Fig. 2 [9]. During the last decade, ML, and DL has demonstrated promising contributions to many research and engineering areas, such as data mining [10], medical imaging [11], communication [12], multimedia [13], geoscience [14], remote sensing classification [15], real-time object tracking [16], computer vision-based fault detection [17], and so forth. The integration of advanced information and communication technologies, specifically Internet of Things (IoT), in the power grid infrastructures, is one of the main steps towards the smart grid. Since the vital capability of IoT devices is their capability to communicate data to other devices in a more pervasive fashion, and hence a massive amount of data is made available at the control centers. Such meaningfully enhanced system condition awareness and data availability demands for ML-based solutions and tools to conduct efficient data processing and analysis, to encourage the system operational management and decision-making [18]. Therefore, ML has been applied in various fields of power system, such as load forecasting [19], fault diagnosis [20], substation monitoring [21], reactive power control [22], unit commitment [23], maintenance scheduling [24], wind power prediction [25], energy management [26], load restoration [27], solar power prediction [28], state estimation [29], TSA [30], economic dispatch [31], and electricity price forecasting [32].

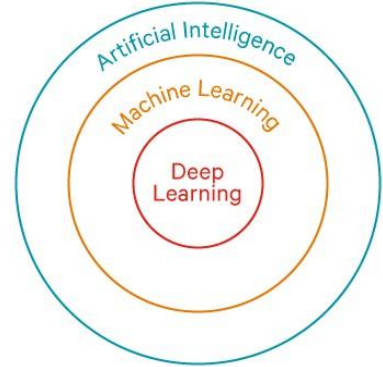

Figure 1.  ML as a subfield of artificial intelligence

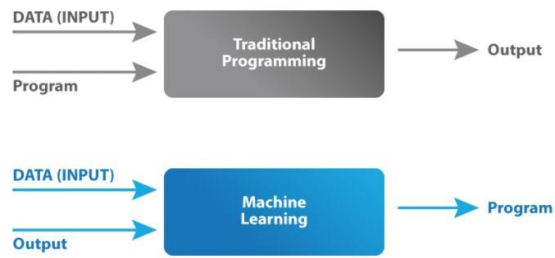

Figure 2.  Traditional programming vs. ML

## III. CLASSIFICATION OF MACHINE LEARNING

ML is generally classified into three broad types [18], as shown in Fig. 3. A brief description of each type is given below.

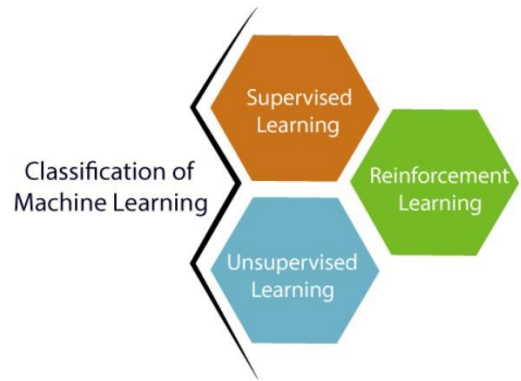

Figure 3.  Types of ML

*1)* Supervised Learning (SL)

In supervised ML (SML), the aim is to learn a mapping between the inputs to outputs based on a given labeled set of input/output pairs in the training set. In this kind of learning, each example is a pair consisting of an input object (normally a vector) and a desired output value. A supervised learning (SL) algorithm examines the training data and generates an inferred function, which can be applied for mapping new examples. Some common SL algorithms neural network ANN, SVM, DTs, Naïve Bayes, and K-Nearest Neighbor (KNN) [33]. The generic framework for SL is illustrated in Fig. 4.

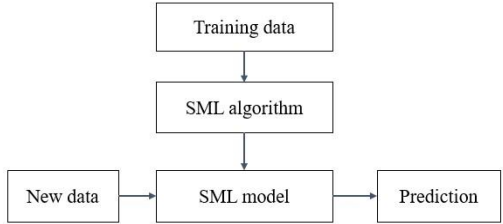

Figure 4. SL generic framework

*2) Unsupervised Learning (UL)*

In unsupervised ML (UML), the training of an algorithm is performed, using information that is neither labeled nor classified, such that the algorithm may cluster the information based on similarity or difference. In contrast to the SL that makes use of labeled data, UL allows to incorporate probability densities over inputs.

The goal of UL is to discover hidden patterns in unlabeled data. Some of the most common algorithms used in UL include clustering and anomaly detection [33]. The generic framework for UL is illustrated in Fig. 5.

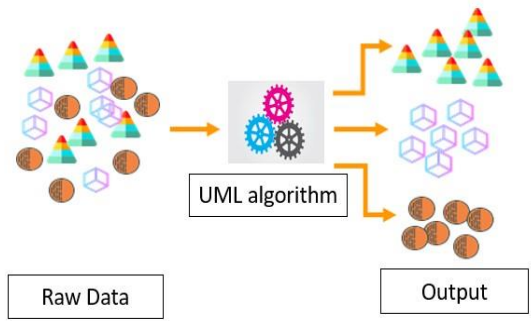

Figure 5. UL generic framework

*3) Reinforcement Learning (RL)*

Reinforcement learning (RL) is an iterative process to foretell the next optimal step to perform a task to get a final reward. In each stage, the DL agent receives an award when it moves in the direction of the goal. RL is appropriate for training a computer to drive a vehicle or playing a game against an opponent [34]. Basically, in RL, an agent interacts with its environment and adapts its actions, based on the reward received in response to its actions [35]. RL differs from SL in the sense that it does not need labelled input/output pairs be presented and does not need sub-optimal actions to be unambiguously adjusted. Instead, the emphasis is on finding a balance between exploration (of uncharted territory) and exploitation (of current knowledge) [36].

In RL process, the environment gives the agent a state. The agent chooses an action and receives a reward from the environment along with the new state. This learning process persists, until the goal is accomplished. The generic framework for RL is illustrated in Fig. 6. Commonly used SMLs include ANN, SVM, DTs, random forest, and Naïve Bayes; however, the focus of the present work is on ANN-based SML algorithm.

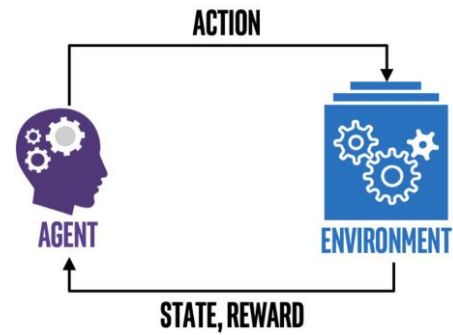

Figure 6. RL generic framework

IV. ARTIFICIAL NEURAL NETWORK: BACKGROUND AND OVERVIEW

The development of ANNs was inspired by the studies of the central nervous system of the human, where the nodes and the interaction within themselves, are to mimic the brains neurons and their synaptic connections. By introducing a training data set to the network, the synaptic weights are iteratively strengthened, until the response of the network follows the output data, like the learning process in the biological brain [37-38]. ANNs are powerful processing tools, enfolding the ability of learning from experience. From a general viewpoint, ANNs are a data-driven, black box technique, aiming at learning and modeling the input-output relationship, of a given process, from the knowledge of a set of input-output measurements only. ANNs have been applied, with remarkable performances, in various black box modeling tasks, involving classification [39-41] and function approximation [42-44]. Generally, ANNs have three layers: input, hidden, and output. The input layer contains the initial data which is fed into the neural network; the output layer produces the results for the given inputs, and the hidden layer is an intermediate layer between input and output layer, where all the required computation is done, i.e., the hidden layer performs nonlinear transformations of the inputs entering the network [42-43].

V. COMPONENTS OF ARTIFICIAL NEURAL NETWORK

This section will briefly describe various components of a typical ANN [45-46].

*1) Neurons*

ANNs are comprised of artificial neurons which are theoretically derived from biological neurons.

Each artificial neuron has inputs and produces a single output, which can be directed to numerous other neurons. The inputs can be the feature values of a sample of external data, such as images or documents, or they can be the outputs of other neurons. The outputs of the final output neurons of the neural net achieve the task, such as image recognition. To determine the output of the neuron, the weighted sum of all the inputs is computed, weighted by the weights of the connections from the inputs to the neuron. Consequently, a bias term is added to this sum. This weighted sum is occasionally called the activation. This weighted sum is then passed through a (usually nonlinear) activation function for output generation.

*2)* Connections and Weights

"The network consists of connections, each connection providing the output of one neuron as an input to another neuron. Each connection is assigned a weight that represents its relative importance. A given neuron can have multiple input and output connections."

*3)* Activation Function

Activation functions are functions used in neural networks to compute the weighted sum of input and biases, which is used to decide whether a neuron can be fired or not. Activation function can be either linear or non-linear depending on the function it represents.

*4)* Layers

"The neurons are typically organized into multiple layers, especially in DL. Neurons of one layer connect only to neurons of the immediately preceding and immediately following layers. The layer that receives external data is the input layer. The layer that produces the ultimate result is the output layer. In between them are zero or more hidden layers. Single layer and unlayered networks are also used. Between two layers, multiple connection patterns are possible. They can be fully connected, with every neuron in one layer connecting to every neuron in the next layer. They can be pooling, where a group of neurons in one layer connect to a single neuron in the next layer, thereby reducing the number of neurons in that layer. Neurons with only such connections form a directed acyclic graph and are known as feedforward networks. Alternatively, networks that allow connections between neurons in the same or previous layers are known as recurrent networks."

*5)* Hyperparameter

A hyperparameter is a constant parameter whose value is set before the learning process begins. The values of parameters are determined using the process of learning. Examples of hyperparameters include number of neurons, the number of hidden layers and batch size.

*6)* Loss Function

The loss function (or a cost function) is one of the most significant component of the ANN. It essentially represents the prediction error of neural network, and the approach to compute the loss is known as the loss function. The loss function simply computes the absolute difference between the predicted and the actual value.

To the best of author's knowledge, there exists no work which comprehensively reviews the research work related to ANN for TSA of power systems. Thus, the main objective and contribution of the current paper is to review major works related to ANN for TSA of power systems and provide research gaps and recommendations for future work.

## VI.   LITERATURE REVIEW: APPLICATION OF ANN FOR TSA

Application of ANN to power system is an area of soaring interest; the main reason being the ability of ANN to process and learn intricate nonlinear relations [47]. Moreover, they possess the ability of parallel processing of data. In ANN-based TSA, a relation mapping is established between the input features and the output results of a stability assessment, based on many offline simulations. Thus, they have been widely applied to create this relation mapping by numerous research [48, 49-51].

Reference [48] used ANNs to predict critical clearing time (CCT) for a small test power system. Reference [49] used an individual TEF approach to predict energy margin and stability. Reference [50] devised an integrated approach of unsupervised and SL for TSA. Reference [52] proposed a fast pattern recognition and classification method for states of dynamic security. In [53], ANNs were used to predict stability of a system consisting of 227 buses and 53 generators. Reference [54] applied the recurrent Radial Basis Function (RBF) and the Multi-Layer Perceptron Neural Network (MLPNN) for predicting rotor angles and angular velocities of synchronous machines. Reference [55] used ANN to classify system stability status for various contingencies. In [56], the nonlinear mapping relation between the transient energy margin and the generator power, at different fault clearing time (FCT), was established by using the multilayer Feedforward Neural Network (FNN). Lyapunov's direct method, based on the system dynamic equivalents, was used as a fast method to obtain the training set for the ANN. Reference [57] presented a novel ANN-based global online fault detection, pattern classification, and relaying detection scheme, for synchronous generators (SGs) in interconnected electric utility networks. The online ANN based relaying scheme classified fault existence and fault type as either transient stability or loss of excitation, and the allowable CCT, and loss of excitation type as either open circuit or short circuit condition.

An innovative two-layer, fuzzy hyperrectangular composite neural network was proposed, in [58], to determine real-time transient stability prediction. In [59], investigation was conducted for enhancing transient stability, by applying auxiliary controls for controlling power flow of High Voltage Direct Current (HVDC). The current controller model and the line dynamics were integrated in the stability analysis. A multi-machine system with a neural network controller was established to boost the stability of the system. Reference [60] discussed the issue of ANN input dimension reduction. Two different methods, for TSA application, were discussed and compared for efficiency and accuracy. Reference [61] described a neural network-based, adaptive pattern recognition approach, for estimation of the CCT. Reference [62] proposed an application of ANN, for contingency screening and ranking of a power system, with respect to transient stability. Reference [63] suggested a method of TSA, by adaptive pattern recognition which makes use of an ANN. Reference [64] aimed to examine the use of ANNs, in the analysis of the transient stability of a power system (determination of CCT for short-circuit faults type, with transmission line outage), using a supervised FNN.

In [65], a multilayer feedforward ANN is employed for the online TSA of a power system. Reference [66] used RBF Neural Network (RBFNN) as a control scheme, for the Unified Power Flow Controller (UPFC), to improve the transient stability performance of a multimachine power system. Reference [67] focused on validating the accuracy of ANN for evaluating the transient stability of a single machine infinite bus system. The fault CCT, obtained through ANN, was compared with the results, obtained through the traditional Equal Area Criterion (EAC) method. The multilayer FNN concept was applied to the test system. Reference [68] presented a comparative analysis of two different ML algorithms, i.e., ANN and SVM, for online transient stability prediction, considering various uncertainties (load, network topology, fault type, fault location, and fault clearing time). The results showed that the performance of ANN was way better than that of SVM as its classification metrics and computational time was determined to be superior.

Reference [69] proposed ANN-based supervised ML, for predicting the transient stability of a power system, considering uncertainties of load, faulted line, fault type, fault location, and fault clearing time. The training of the neural network was achieved using appropriate system features as inputs, and probabilistic transient stability (PTS) status indicator as the output. Reference [70] investigated the framework for PTS in power systems and the application of ANN to improve its assessment process. Numerous uncertain factors such as faulted line, fault type, fault location, and fault clearing time were part of the analysis. The results obtained indicated the effectiveness of the suggested algorithm such that it can be applied to predict transient stability of any large-scale practical power system.

In [71], a comparative analysis between DT, SVM and ANN for two datasets demonstrated that TSA using ML is system specific. It was also shown that he performance between the two sets of algorithms fluctuates considerably as the parameters of the network shift. A TSA and instability mode assessment approach based on convolutional neural network was presented in [72]. The technique takes the bus voltage phasor sampled by Phasor Measurement Units (PMUs) during a short observation window after disturbance as input, and outputs the forecast result swiftly in terms of stability, aperiodic instability, or oscillatory instability. In [73], a novel hybrid intelligent system was devised for predicting transient stability. The system was composed of a preprocessor, an array of neural networks and an interpreter. The preprocessor partitioned the whole set of synchronous machines into subsets, each one including two generators.

A novel method for power system TSA was suggested in [74] based on voltage phasor and CNN (Convolution Neural Network). Firstly, using the DL technique, a dynamic display of power system transient process in the voltage phasor complex plane was assembled. Secondly, based on CNN and the image of voltage phasor complex plane, the power system transient stability fast estimation prototype was suggested. A direct method based on Type-2 fuzzy neural network for TSA was suggested in [75]. The Type-2 fuzzy logic had the ability to tackle the uncertainty in the measurement of power system parameters. On the contrary, a multilayer perceptron (MLP) neural network possesses expert knowledge and ability to learn. The devised hybrid approach combined both of these capabilities to attain a precise estimation of CCT, which is an index of TSA.

Reference [76] proposed TSA of a large 87-bus system using a unique approach known as the Probabilistic Neural Network (PNN) by integrating feature selection and extraction approaches. Transient stability was predicted based on the generator relative rotor angles obtained from time domain simulations. It was concluded that the PNN with the incorporation of feature reduction approaches reduced the PNN training time without impacting the correctness of the classification results. Real-time TSA was presented in a data driven framework in [77], by incorporating the temporal relations of the predictors using Recurrent Neural Network (RNN) with Long Short-Term Memory (LSTM) units. The presented method was illustrated on the IEEE 39-bus test system and produced extraordinary test results compared with a SVM benchmark.

In [78], a TSA system based on the long short-term memory network was developed. By suggesting a temporal self-adaptive scheme, the presented network aimed to balance the trade-off between evaluation accuracy and response time. Case studies on three power systems demonstrated the effectiveness of the suggested TSA approach. Reference [79] presented a methodical approach for building and renovating an exact transient stability classifier. Firstly, the time-series trajectories of generators after disturbance were used as the inputs, and consequently, a CNN ensemble method was suggested to generate the transient stability predictor using these multi-dimensional data. The simulation results of two power systems demonstrated the efficacy of the presented technique. Reference [80] explained the capability of ANN for predicting the CCT of power system. The training of ANN was done using selected features as input and CCT as desire target. A single contingency was used and the target CCT was determined using time domain simulation. The simulation showed that ANN can deliver rapid and accurate mapping which makes it suitable for online applications.

Reference [81] presented an application of ANN for monitoring TSA incorporating system topology changes. Offline trained ANN was used in indicating the appropriate remedial action for online operation to neutralize the system instability. The simulation results indicated that the ANN is a useful tool to assess online transient stability with acceptable accuracy. In [82], the Gated Graph Neural Network (GGNN) was used to predict the transient stability and infer the kind of disturbance leading to the instability of power system. Firstly, Conditional Generative Adversarial Network (CGAN) was applied to generate unstable samples. Consequently, the real-time data was input into the trained TSA model and the transient stability of power system was assessed. Reference [83] applied Graph Neural Network (GNN) to include topology information to the model and consequently, realized the combination of electrical information and network topology information to establish a transient stability evaluation model. A simulation example verified the feasibility of the model in TSA and proved that the model can generalize to any grid topology.

A method based on convolutional neural network was devised for power TSA in [84], which can overwhelm the flaws of conventional assessment methods and fulfill the obligations with high evaluation precision of TSA issues. The training methods of convolutional neural network were optimized according to the characteristics of TSA problem, and the batch normalization algorithm was added to establish the TSA model. Reference [85] termed a neural network based adaptive pattern recognition approach by making a thorough analysis on a power system for estimation of the CCT. Back propagation technique was used to adjust the weights. Analytical calculations were compared with the values obtained by neural network.

In [86], a unique approach for the prediction of online transient stability margin was presented. Combined with a Geographic Information System (GIS) and transformation rules, the topology information and pre-fault power flow characteristics were extracted by 2D computer-vision-based power flow images. Also, a CNN-based comprehensive network was established to map the relationship between the steady-state power flow and the generator stability indices under the expected set of contingencies. The presented approach can be used to evaluate the transient stability margin rapidly and quantitatively, and the simulation results validated the efficacy of the suggested method. In [87], Recurrent Graph Convolutional Network (RGCN) based multi-task TSA, was presented. Both the Graph Convolutional Network (GCN) and the LSTM unit were aggregated to form the RGCN. Test results on IEEE 39 Bus system and IEEE 300 Bus system indicated the dominance and robustness of the suggested method over existing models under various scenarios. Some other significant work, associated with ANN application to TSA, can be found in [88-95].

## VII. Research Gaps and Future Recommendations

Based on the detailed literature review and to the best of author's knowledge, there exists no research works on PTS, which specifically uses ANN-based ML approach, considering all the uncertainties of load, faulted line, fault type, fault location (on the line), and FCT. Although, [96] used ANN for PTS, but the approach only considers the uncertainty of load (ignoring other uncertainties of fault type, fault location, and FCT). Moreover, [89] specifically mentions the potential of ANN for online Dynamic Security Assessment (DSA). In addition, [97-103] strongly indicate that ML is a favorable and forthcoming approach for online DSA (which also includes TSA). Thus, one of the main research gap is to predict PTS status using an ANN-based ML approach.

Also, there is a dearth of work which incorporates renewable energy in TSA prediction using ANN. As the amount of renewable energy is increasing continually in the power system, the dynamics of power system are becoming more intricate. More comprehensive studies and simulations are required to understand the behavior of the system under renewable energy integration. Network topology changes are often ignored in the existing research work. It is significant to train the ANN model for changing topologies. Novel techniques must be researched and applied such that the ANN model can adapt to any topology of the power system. Moreover, a majority of research articles assume that PMU online data is comprehensive and error-free; however, in real life scenarios, the obtained value of a parameter may be inaccurate or unavailable due to jamming, malfunctioning, or even cyber-attacks [104].

Currently, ANN-based TSA methods face some challenges. First, it is very difficult to obtain large-scale, balanced data of random input variables with accurate labels in real-world scenarios. On top of that, existing ANN-based TSA methods act as a black box which have poor interpretability, which also limits their application in actual power systems [105]. Approaches can be formulated to integrate various kinds of stability into a single index and consequently, utilize ANN to predict its value. It is highly recommended that future research work on ANN based approach to power system transient stability should focus on comprehensive validation of the approach using large scale test system which have similar attributes of uncertainty and randomness as that of a modern power system.

The present study provided a review of some major research works and potential future research avenues associated with ANN application to TSA. This can be a remarkable starting point for researchers in the domain of ML, power system stability and security, particularly in the presence of uncertainty. Recent research [106-117] reveals that there is a lot of scope in this domain, and its potential must be fully investigated.

## VIII. Conclusion and Future Work

Rapid TSA is a critical requirement for correct and timely operation of an electric power system. The traditional methods fail to fulfil this requirement. In this regard, ML approaches can play a significant role. Therefore, this paper provided a review of works related to application of ANN to TSA. It is believed that this review will provide a good basis for researchers in the field of ML and power system stability, and consequently, help them understand the current research status and existing challenges. As a future work, various reviews and case studies can be conducted using other ML approaches and a comparative analysis can be drawn. Moreover, techniques for feature reduction must be devised for large-scale power systems to further reduce the computational efforts. ML can also be integrated with edge computing for modern power systems which will allow the smart devices (smart meters, PMUs, digital relays, etc.) to process the data locally, thereby reducing the dependence on cloud networks.